%Paper: hep-th/9505119
%From: George Papadopoulos <G.Papadopoulos@damtp.cam.ac.uk>
%Date: Fri, 19 May 95 10:46:25 BST


\input phyzzx
\def\cL{{\cal L}}
\def\dplus{=\hskip-5pt \raise 0.7pt\hbox{${}_\vert$} ^{\phantom 7}}
\def\dplusup{=\hskip-5.1pt \raise 5.4pt\hbox{${}_\vert$} ^{\phantom 7}}

%%%%%%%%%%%%%%%%%%%%%%%%%%%%%%%%%%%%%%%%%%%%%%%%%%%%%%%%%%%%%%%%%%
\def\dplus{=\hskip-4.8pt \raise 0.7pt\hbox{${}_\vert$} ^{\phantom 7}}

\def\pmb#1{\setbox0=\hbox{#1} \kern-.025em\copy0\kern-\wd0
\kern0.05em\copy0\kern-\wd0 \kern-.025em\raise.0433em\box0}

\def\cM{{\cal M}}
\def\cL{{\cal L}}
%%%%%%%%%%%%%%%%%%%%%%%%%%%%%%%%%%%%%%%%%%%%%%%%%%%%%%%%%%%%%%%%%%%%

\REF\hpa {P.S. Howe and G. Papadopoulos, Nucl .Phys. {\bf B289} (1987) 264;
Class. Quantum Grav. {\bf 5} (1988) 1647.}
\REF\ghr{S.J. Gates, C.M. Hull and M. Ro{\v c}ek, Nucl. Phys. {\bf B248}
(1984) 157; C.M. Hull, {\it Super Field Theories} ed H. Lee and G. Kunstatter
(New York: Plenum) (1986).}
\REF\hpb{P.S. Howe and G. Papadopoulos, Nucl .Phys. {\bf B381} (1992) 360.}
\REF\chs{C.G. Callan, J.A. Harvey and A. Strominger, Nucl. Phys. {\bf B359}
(1991) 611.}
\REF\vp {Ph. Spindel, A. Sevrin, W. Troost and A. Van Proeyen, Nucl. Phys. {\bf
B308} (1988) 662;
{\bf B311} (1988) 465.}
\REF\dv{F. Delduc and G. Valent, Class. Quantum Grav. {\bf 10} (1993) 1201.}
\REF\bv{G. Bonneau and G. Valent, {\sl Local heterotic geometry in holomorphic
co-ordinates},
hep-th/9401003.}
\REF\gh {G.W. Gibbons and S.W. Hawking, Phys. Lett.  {\bf 78B} (1978) 430.}
\REF\gwg{G.W. Gibbons and P. J. Ruback, Commun. Math. Phys. {\bf 115} (1988)
267.}
\REF\ghb {G.W. Gibbons and S.W. Hawking, Commun. Math. Phys. {\bf 66} (1979)
291.}
\REF\leb{C. LeBrun, J. Differential Geom. {\bf 34} (1991) 223.}
\REF\br{T.H. Buscher, Phys. Lett. {\bf 194B} (1987) 51; {\bf 201B} (1988) 466.}
\REF\rg{A. Giveon and M. Ro{\v c}ek, {\sl Introduction to Duality},
hep-th/9406178;\break
E. \'Alvarez, L. \'Alvarez-Gaum\'e, Y. Lozano, {\sl An introduction to
T-duality in string theory},
hep-th/9410327.}

%%%%%%%%%%%%%%%%%%%%%%%%%%%%%%%%%%%%TITLE
%%PAGE%%%%%%%%%%%%%%%%%%%%%%%%%%%%%%%%%%%%%%%%%%%

\Pubnum{ \vbox{ \hbox{R/95/15} } }
\pubtype{}
\date{May, 1995}

\titlepage

\title{Elliptic monopoles and (4,0)-supersymmetric sigma models with torsion}
\author{ G. Papadopoulos}
%\andauthor{}
\address{D.A.M.T.P
 \break University of Cambridge\break
         Silver Street \break Cambridge CB3 9EW}

%\address{}

\abstract {
We explicitly construct the metric and torsion couplings of two-dimensional
(4,0)-super\-sym\-metric
sigma models with target space a four-manifold that are invariant under a
$U(1)$ symmetry
generated by a tri-holomorphic Killing vector field that leaves in addition the
torsion invariant.
We show that the metric couplings arise from magnetic monopoles on the
three-sphere which is
the space of orbits of the group action generated by the tri-holomorphic
Killing vector field on the
sigma model target manifold. We also examine the global structure of a subclass
of these metrics
that are in addition $SO(3)$-invariant and find that the only non-singular one,
for models with
non-zero torsion, is that of $SU(2)\times U(1)$ WZW model. }
\vskip 1cm

%\centerline{}

\endpage

\pagenumber=2

%%%%%%%%%%%%%%%%%%%%%%%%%%%%%%%%%%%%MACROS%%%%%%%%%%%%%%%%%%%%%%%%%%%%%%%%%%%%%%%%%%

%\font\mybb=msbm10 at 12pt
%\def\bb#1{\hbox{\mybb#1}}
%\def\RN {\bb{R}}
%\def\CN {\bb{C}}

\def\cN {{\cal {N}}}

\def\fff{\vrule width0.5pt height5pt depth1pt}
\def\pp{{{ =\hskip-3.75pt{\fff}}\hskip3.75pt }}

\def\cM {{\cal{M}}}

%\sequentialequations

%%%%%%%%%%%%%%%%%%%%%%%%%%%%%%%INTRODUCTION%%%%%%%%%%%%%%%%%%%%%%%%%%%%%%%%

\chapter{Introduction}

It has been known for sometime that there is an interplay between the number of
supersymmetries which
leave the action of a sigma model invariant and the geometry of its target
space.  More
recently, sigma models with symmetries generated by Killing vector fields are a
fertile area for
investigation of the properties of T-duality. The couplings of two-dimensional
sigma models
are the metric $g$ and  a locally defined two-form $b$ on the sigma model
manifold $\cM$. The closed
three-form $H={3\over2} db$  is called Wess-Zumino term or torsion\foot{In
two-dimensions, the sigma
model action can be extended to include terms with couplings other than $g$ and
$b$. For
example, the fermionic sector in (p,0)-supersymmetric sigma models has as a
coupling a Yang-Mills
connection on $\cM$.}.  A special class of supersymmetric sigma models are
those with (4,0)
supersymmetry [\hpa, \ghr]. These models are ultra-violet finite [\hpa, \hpb]
and arise naturally in
heterotic string compactifications [\chs]. It is known that the
target space of (4,0)-supersymmetric sigma models admits three complex
structures that obey the
algebra of imaginary unit quaternions. When the torsion $H$ vanishes  the
target space is
hyper-K\"ahler with respect to the metric $g$.  In the presence of torsion, the
geometry of the
target space of (4,0)-supersymmetric sigma models may not be hyper-K\"ahler and
new geometry arises.
It can be shown that given a (4,0)-supersymmetric sigma model with target space
a four-dimensional
hyper-K\"ahler manifold with metric $g$ ($H=0$), it is possible to construct
another one with metric
$g^F=e^F g$ and  torsion  $H={}^*dF$ provided that
$F$ is a harmonic {\it function} with respect to the metric $g$ [\chs]. The
converse though it is not
necessarily true.  For example the bosonic Wess-Zumino-Witten (WZW) model with
target space
$SU(2)\times U(1)$ admits a (4,0)-supersymmetric extension [\vp] but its target
manifold is not
hyper-K\"ahler as it can be easily seen by observing that the second deRham
cohomology group
of $SU(2)\times U(1)$ vanishes. To construct (4,0)-supersymmetric sigma models
with torsion that
are not conformally related to hyper-K\"ahler ones, the authors of ref. [\dv]
employed harmonic
superspace methods and found a generalisation of the Eguchi-Hanson and Taub-NUT
geometries that have
{\sl non-zero} torsion.  Subsequently, it was shown in [\bv] that the above
Eguchi-Hanson geometry
with torsion is conformally equivalent to the standard Eguchi-Hanson one.

One result of this letter is to give the most general metric and torsion
couplings of a
(4,0)-supersymmetric sigma model with target space  a four-dimensional manifold
$\cM$ that admits a
tri-holomorphic Killing vector field $X$ that in addition leaves the torsion
$H$ invariant.  Let
${\cal N}$ be the space of orbits of the $U(1)$ group action generated by the
vector field $X$ on
the sigma model target manifold $\cM$ (away from the fixed points of $X$).  We
shall show that the
non-vanishing components of the metric $g$ and torsion $H$ are given as
follows:
$$\eqalign{
ds^2&=V^{-1} (dx^0+\omega_i dx^i)^2+ V \gamma_{ij} dx^i dx^j
\cr
H_{ijk}&=\lambda \epsilon_{ijk} V,\qquad \qquad i,j,k=1,2,3\ ,}
\eqn\pone
$$
where  $\omega$ is a one-form, $V$ is a scalar function and $\gamma$ is a
three-metric on ${\cal N}$, so all
depend on the co-ordinates $\{x^i; i=1,2,3\}$ of $\cN$, and $\lambda$ is a real
constant. The
tensors $\gamma$, $V$ and $\omega$ satisfy the following conditions:
$$
\eqalign{
 2\partial_{[i} \omega_{j]}&=\epsilon_{ij}{}^k \partial_k V
\cr
{}^{(3)}R_{ijkl}&=2 \lambda^2 \gamma_{k[i} \gamma_{j]l}\ ,}
\eqn\ptwo
$$
where $\epsilon_{ijk}$ is adopted to the metric $\gamma$.
The  scalar function $V$ is therefore a harmonic function of ${\cal N}$ with
respect to the
metric $\gamma$ and the metric $\gamma$ has {\sl positive} constant curvature.
The latter implies that the three-manifold ${\cal N}$ is elliptic and therefore
locally
isometric to the (round) three-sphere.   In the case that the constant
$\lambda$ is zero one recovers
the Gibbons-Hawking metrics [\gh, \gwg] that are associated with monopoles on
the flat
(${}^{(3)}R=0$) three-space ${\cal N}={\cal R}^3$.  In the case that
$\lambda\not=0$ and $V$
constant, we recover the (4,0)-supersymmetric WZW model with target space the
manifold $SU(2)\times
U(1)$.  We shall also study the global properties of a subclass of these
metrics that are
invariant under an isometry group with Lie algebra $u(2)$ acting on $\cM$ with
three-dimensional
orbits ($u(2)$-invariant metrics), for $\lambda\not=0$, and we shall find that
apart from the metric
of $SU(2)\times U(1)$ WZW model all the remaining ones are {\sl singular}.

The material of this letter is organised as follows: In section two, the
geometry of
(4,0)-supersymmetric sigma models is briefly reviewed.  In section three, the
derivation of equations
\pone\ and \ptwo\ is presented. In section four, the singularity structure of a
subclass of the
metrics \pone\ that are $u(2)$-invariant is investigated.  Finally in section
five we give our
conclusions.

%%%%%%%%%%%%%%%%%%%%%%%%chapter2%%%%%%%%%%%%%%%%%%%%%%%%%%%%%%%%%%%%%%%%%%%%%%%%%

\chapter{Geometry and (4,0) supersymmetry}

Let $\cM$ be a Riemannian manifold with metric $g$ and a locally defined
two-form $b$. The action of
the (1,0)-supersymmetric sigma model with target space $\cM$ is
$$
I=-i\int\! d^2 x d \theta ^+  (g_{\mu\nu} +b_{\mu\nu})
D_+\phi^\mu\partial_=\phi^\nu  \ ,
\eqn\aone
$$
where  $(x^\pp,x^=, \theta^+)$ are the co-ordinates of (1,0) superspace
$\Xi^{(1,0)}$ and $D_+$ is
the  supersymmetry derivative
($D_+^2=i \partial_\pp$); $(x^\pp,x^=)=(x+t, t-x)$ are light-cone co-ordinates.
 The fields $\phi$
of the sigma model are maps  from the (1,0) superspace, $\Xi^{(1,0)}$, into the
target manifold
${\cal M}$.

To construct sigma models with (4,0) supersymmetry, we introduce the
transformations
$$
\delta_I\phi^\mu=a_{-r} I_r{}^\mu{}_\nu D_+\phi^\nu
\eqn\atwo
$$
written in terms of (1,0) superfields, where $I_r$, r=1,2,3, are  (1,1) tensors
on $\cM$ and
$a_{-r}$ are the parameters of the transformations.  The commutator of these
transformations
closes to translations\foot{Due to the classical superconformal invariance of
the model the
parameters $a_{-r}$ of the supersymmetry transformations can be chosen to be
semi-local in which
case the commutator of supersymmetry transformations closes to translations and
supersymmetry
transformations.}   provided that
$$
\eqalign{
I_r I_s&=-\delta_{rs}+\epsilon_{rst} I_t
\cr
N(I_r)^\mu{}_{\nu\rho}&= 0 \ ,}
\eqn\athree
$$
where
$$
N(I_r)^\mu{}_{\nu\rho}=I_r{}^\kappa{}_\nu\partial_\kappa
I_r{}^\mu{}_\rho-I_r{}^\mu{}_\kappa\partial_\nu
I_r{}^\kappa{}_\rho-(\rho\leftrightarrow \nu)
\eqn\afour
$$
is the Nijenhuis tensor of $I_r$. The conditions \athree\ imply that  $I_r$,
$r=1,2,3$, are complex
structures that satisfy the algebra of imaginary unit quaternions.

The action \aone\ of (1,0)-supersymmetric sigma model is invariant under the
(4,0) supersymmetry
transformations \atwo\ provided that, in addition to the conditions obtained
above for the
closure of the algebra of these transformations, the following conditions are
satisfied:
$$
 g_{\kappa(\mu} I_r{}^\kappa{}_{\nu)}=0\ , \qquad \nabla^{(+)}_\mu
I_r{}^\nu{}_\rho=0\ ,
\eqn\afive
$$
where $\nabla^{(\pm)}$ is the covariant derivative of the connection
$$
\Gamma^{(\pm)}{}^\mu_{\nu\rho}=\{{}^\mu_{\nu\rho}\}\pm H^\mu{}_{\nu\rho}\ ,
\eqn\asix
$$
and
$$
H_{\mu\nu\rho}={3\over 2} \partial_{[\mu} b_{\nu\rho]}\ .
\eqn\aseven
$$
Note that if $I_r$ are integrable, $N(I_r)=0$, and covariantly constant with
respect to the
$\nabla^{(+)}$ covariant derivative, $\nabla^{(+)} I_r=0$, then the torsion $H$
is (2,1)-and
(1,2)-form with respect to all complex structures $I_r$.  Note also that, if
the torsion $H$ is zero,
the above conditions simply imply that the sigma model target manifold is
hyper-K\"ahler with
respect to the metric $g$

Next consider the transformations
$$
\delta\phi^\mu=\epsilon^a X_a^\mu(\phi)\ ,
\eqn\aeight
$$
of the sigma model field $\phi$, where $\{\epsilon^a; a=1,2, \dots\}$ are the
parameters of
these transformations and $\{X_a;; a=1,2, \dots\}$ are vector fields on the
sigma model manifold
$\cM$.  The action \aone\ is invariant under these transformations up to
surface terms provided that
$$
\nabla_\mu X_{a\nu}+\nabla_\nu X_{a\mu}=0, \qquad X_a^\kappa
H_{\kappa\mu\nu}=\partial_{[\mu}
u_{a\nu]} \ .
\eqn\anine
$$
These conditions imply that $\{X_a;; a=1,2, \dots\}$ are Killing vector fields
on the sigma model
manifold $\cM$ and leave the closed three-form $H$ invariant .
The commutator of the \aeight\ transformations with the (4,0)
supersymmetry transformations is
$$
[\delta_\epsilon, \delta_I]\phi^\mu=\epsilon^b a_{-r} \cL_{X_b} I_r{}^\mu{}_\nu
D_+\phi^\nu
\eqn\aten
$$
This commutator closes on the existing symmetries of the theory, if we take
$$
\cL_{X_a} I_r{}^\mu{}_\nu=0\ ,
\eqn\aeleven
$$
where $\cL_{X_a}$ is the Lie derivative with respect to the vector field $X_a$.
All three complex structures $I_r$ are invariant under the vector field $X_a$,
{\it i.e.} $X_a$
is tri-holomorphic. Note that it is possible to relax the above condition. For
example, we can take
that the complex structures $I_r$ rotate under the isometries but this
possibility will not be
considered here.  Finally the commutator of the transformations  \aeight\ with
themselves closes
provided that
$$
[X_a, X_b]=f_{ab}{}^c X_c
\eqn\atwelve
$$
where $f_{ab}{}^c$ are the structure constants of a Lie algebra.

%%%%%%%%%%%%%%%%%%%%%%%%%%%%%%%%CHAPTER3%%%%%%%%%%%%%%%%%%%%%%%%%%%%

\chapter{(4,0) supersymmetry and four-dimensional geometry}

To find the geometry of the target space of sigma models with (4,0)
supersymmetry, one has to solve
the conditions \athree\ and \afive\ of the previous section.  For this we will
restrict ourselves to
four-dimensional target spaces $\cM$ and  we will assume that there is a
Killing  vector
field $X$ on $\cM$ that leaves the torsion $H$ and the complex structures $I_r$
invariant. As we
have seen in the previous section, these conditions on the vector field $X$ are
those
required for the invariance of the action \aone\ under the transformations
\aeight\ of the sigma
model fields $\phi$ and the closure of the commutator these transformations
with the (4,0)
supersymmetry transformations.

Next we adopt coordinates  $\{x^0, x^i; i=1,2,3\}$ on the sigma model manifold
along the Killing
vector field $X$, {\it i.e.}
$$
X={\partial\over\partial x^0}\ .
\eqn\afourteen
$$
Then the metric $g$ and the torsion $H$ are written as follows:
$$
ds^2= V^{-1} (dx^0+\omega_i dx^i)^2+ V\gamma_{ij} dx^i dx^j\ ,
\eqn\atwelve
$$
and
$$
H_{0ij}=\partial_{[i}u_{j]}, \qquad H_{ijk}=\epsilon_{ijk} U\ ,
\eqn\athirteen
$$
where $\gamma$, $u$, $V$ and $U$ are tensors of the space of orbits, ${\cal
N}$, of the group action
generated by the Killing vector field $X$ on $\cM$ (away from the fixed points
of $X$) and depend
only upon the co-ordinates $\{x^i; i=1,2,3\}$.   The tensor
$\gamma$ is a metric on ${\cal N}$ and the tensor $\epsilon_{ijk}$ is adopted
to the
three-metric $\gamma$.

We introduce the frame
$$\eqalign{
e^{\underline {0}}&=V^{-{1\over 2}}(dx^0+\omega_i dx^i)\ ,
\cr
e^{\underline {r}}&= V^{{1\over 2}} E^{\underline {r}}_i dx^i \ ,\qquad
{\underline {r}}=1,2,3    }
\eqn\afifteen
$$
associated with the metric \atwelve, and its dual
$$\eqalign{
e_{\underline {0}}&=V^{1\over 2} \partial_0
\cr
e_{\underline {r}}&= V^{-{1\over 2}} E_{\underline {r}}^i \big(
\partial_i-\omega_i\partial_0\big)\ ,}
\eqn\afifteena
$$
where $\partial_0={\partial\over \partial x^0}$, $\partial_i={\partial\over
\partial x^i}$, and
$E^{\underline {r}}$ and $E_{\underline {r}}$ is a frame of the metric
$\gamma$ and its dual, respectively.  Next we introduce three invariant (1,1)
tensors on $\cM$ as
follows:
$$
I_r= e_{\underline {0}}\otimes e^{\underline {r}}- e_{\underline {r}}\otimes
e^{\underline
{0}}-\epsilon_{rst}  e_{\underline {s}}\otimes  e^{\underline {t}}
\eqn\afifteenb
$$
Contracting with the metric $g$, we can show that
$$
{\cal I}_r\equiv \gamma_{\mu\kappa} I_r{}^\kappa{}_\nu dx^\mu \otimes dx^\nu= 2
e^{\underline 0}
\wedge e^{\underline r}-\epsilon_{rst} e^{\underline s}e^{\underline t}\ .
\eqn\asixteen
$$
are two-forms on $\cM$ and so the first condition of eqn. \afive\ is satisfied.
It is  straightforward to verify that the tensors $I_r$ are almost complex
structures that
satisfy the algebra of imaginary unit quaternions and so  the first condition
of eqn. \athree\
is also satisfied.  Furthermore, it can be shown that $\{I_r; r=1,2,3\}$ (eqn.
\afifteenb) is the
most general set of almost complex structures, up to an $SO(3)$
(gauge) rotation of the frame $E^{\underline {r}}$ of the three-metric
$\gamma$, that obey the
algebra of imaginary unit quaternions and are invariant under the group action
generated by the
Killing vector field $X$.  Therefore $X$ is a tri-holomorphic vector field
and  satisfies eqn. \aeleven.   So it remains to find the conditions on
$V,U,u$ and
$\gamma$ in order the metric \atwelve, antisymmetric tensor
\athirteen\ and the almost complex structures $I_r$  satisfy the second
condition  of \athree\ and
the  second condition of \afive.  Combining the second condition of
\afive\ and \aeleven, we can show that
$$
\nabla^{(+)}_i X_j
\eqn\abone
$$
is a (1,1)-form with respect to all almost complex structures $I_r$ which in
turn implies that
$$
2 \partial_{[i} \omega_{j]}- 2 V \partial_{[i}
u_{j]}=\epsilon_{ij}{}^k\partial_k V\ .
\eqn\abtwo
$$
Next using the fact that the torsion $H$ is (2,1)-and (1,2)-form with respect
to all almost complex
structures $I_r$, the condition
$$
\nabla^{(+)}_{[\mu} {\cal I}^r_{\nu\rho]}=0
\eqn\abthree
$$
implies that
$$
\eqalign{
du&=0, \qquad U V^{-1}=\lambda,
\cr
{}^{(3)}R_{ijkl}&=2\lambda^2 \gamma_{k[i} \gamma_{j]l}\ , }
\eqn\aseventeen
$$
where ${}^{(3)}R$ is the curvature of the metric $\gamma$ and $\lambda$ is a
real constant.  After
some computation, it can be shown that the almost complex structures are
integrable without further
conditions, {\sl i.e.} their Nijenhuis tensor vanishes. Finally,
the second condition of \afive\ follows from the integrability of $I_r$, the
condition
\abthree\ and the fact that $H$ is (2,1)- and (1,2)-form with respect to all
complex
structures $I_r$.  Substituting the equations \abtwo\ and \aseventeen\ back
into \atwelve\ and
\athirteen\ we get the metric $g$ and torsion $H$ of eqn. \pone.

Apart from the special cases for which either $\lambda=0$ or $V$ constant
mentioned in the
introduction,  new metrics can be found by taking
$$
V(x^i)=c_0+\sum^N_{n=1} c_n G(x,x_n)
\eqn\aeighteen
$$
where $\{G(x,x_n); n=1,\dots, N\}$ are the Green's functions of the
Laplace-Beltrami operator $\Delta$ associated with the metric $\gamma$, $\{c_0,
c_n; n=1, \dots,
N\}$ are real constants and $\{x_n; n=1, \dots, N\}$ are $N$ points in ${\cal
N}$.

The metric $g$ of \pone\  is not conformally equivalent to a hyper-K\"ahler one
if $\lambda\not=
0$ and $x^0$ is an angular co-ordinate.
To show this, let
$$
g^F= e^F g
\eqn\atwenty
$$
where $F$ is a function of $\cM$.  Since $I_r$ are complex structures, to prove
that $g^F$ is
hyper-K\"ahler it is enough to show that the three two-forms ${\cal
I}^F_r{}_{\mu\nu} \equiv
g^F_{\mu\kappa} I_r{}^\kappa{}_\nu$ are closed, {\sl i.e.}
$d{\cal I}^F_r=0$.
After some computation, it can be shown that $d{\cal I}^F_r=0$ implies that
$$
\eqalign{
\partial_0 F&=-2 \lambda V^{-1}
\cr
\partial_i F&=-2 \lambda V^{-1} \omega_i\ .}
\eqn\atwentytwo
$$
It is clear that, if $\lambda=0$ (the torsion $H$ is zero), the most general
solution
of \atwentytwo\  is that $F$ is equal to a real constant. In this case
$g$ is hyper-K\"ahler and so is $g^F$.  In the case that $\lambda\not=0$, we
differentiate the second
equation in \atwentytwo\ with respect to $\partial_0$ and we get
$$
\partial_0\partial_i F=0
\eqn\atwentythree
$$
but
$$
\partial_0\partial_i F\equiv \partial_i\partial_0 F=-2 \lambda \partial_i
V^{-1}=0\ ,
\eqn\atwentythreea
$$
which implies that $V$ is constant.  So for the metric $g$ to be conformally
equivalent to a
hyper-K\"ahler metric, $V$ must be a real constant.  However, the case that
$V$ equal to a constant can also be excluded if the co-ordinate $x^0$ is an
angle.  To see this
observe that all solutions $F$ of \atwentytwo\ are linear in $x^0$ and so they
cannot be scalar functions of $\cM$.  For example, the metric $g$ of the
(4,0)-supersymmetric WZW
model with target space the group $SU(2)\times U(1)$ is not conformally
equivalent to a
hyper-K\"ahler one.  Using a similar argument and under the  assumptions that
$\lambda\not=0$
and $x^0$ is an angle, we can also show that the metric $g$ of eqn \pone\ is
not conformally
equivalent to a K\"ahler one.

%%%%%%%%%%%%%%%%%%%%%%%%%%%%%%%%CHAPTER3%%%%%%%%%%%%%%%%%%%%%%%%%%%%

\chapter{$u(2)$-invariant metrics}

The metrics given in  eqn. \pone\ may have singularities. To examine such
global properties of these
metrics, we consider the following example. Let us write the three-metric
$\gamma$ in
spherical polar co-ordinates
$$
d\Omega^2_3=R^2 d\psi^2 +\sin^2\psi d\Omega^2_2
\eqn\gone
$$
where $R$ is the radius of the three-sphere and
$$
d\Omega^2_2=R^2 \big(d\theta^2+\sin^2\theta d\varphi^2\big)
\eqn\gtwo
$$
is the metric on a two-sphere.  Next consider the case that  $V$ is equal to
the Green's function $G$
that depends only on $\psi$.  Such $V$  is
$$
V=c_1 \cot\psi+c_0
\eqn\gthree
$$
where $c_1$ and $c_0$ are real constants.  Then it can be arranged that
$$
\omega= c_1 \cos\theta d\varphi\ .
\eqn\gfour
$$
The metric $g$ with $V$ given in \gthree\ admits an isometry group with Lie
algebra $u(2)$ acting
on $\cM$ with three-dimensional orbits. We will consider three cases the
following.  Case(i) the
constant
$c_0=0$ and the constant $c_1\not=0$, in this case the metric
$g$ exhibits singular behaviour at $\psi=0$ and $\psi={\pi\over 2}$.  The
singularity at $\psi=0$ is
a nut singularity and can be removed by the standard methods of ref. [\ghb]
provided that $0\leq
{x^0\over c_1 R}<4\pi$ ($c_1>0$).  However the singularity at
$\psi={\pi\over2}$ cannot be removed
as it can be easily seen by changing the $\psi$  co-ordinate to
$u=\cot \psi$ and by studying the behaviour of the metric at $u=0$.   Case (ii)
the constants
$c_0\not=0$ and $c_1\not=0$ , in this case it can be arranged by rescaling the
metric to
set $c_0=1$.  The metric $g$ has singular behaviour at
$\psi=0$ and $\cot \psi=-{1\over c_1}$.  The singularity at $\psi=0$ is a nut
singularity and
again it can be removed provided that $0\leq {x^0\over c_1 R}<4\pi$ ($c_1>0$)
as in case (i).  But
the singularity at $\cot \psi=-{1\over c_1}$ cannot be removed as it can be
easily seen by changing
the $\psi$ co-ordinate to $u=\cot \psi+{1\over c_1}$ and by studying the
behaviour of the metric at
$u=0$.  Another way to see that the metric is singular at $\cot \psi=-{1\over
c_1}$ is to observe
that the geodesics of $\cM$ that depend only upon $\psi$ reach the singularity
at finite proper time
and they cannot be extended beyond it. The manifold $\cM$ is therefore
geodesically incomplete.

Next define the new co-ordinate $s=\tan\psi$, then, the metric $g$ is rewritten
as
$$
ds^2=(c_0+c_1 s^{-1})^{-1} \big(dx^0+c_1 \cos\theta d\varphi\big)^2+(c_0+c_1
s^{-1})
\big[{R^2\over 1+s^2} ds^2 +{s^2\over 1+s^2} d\Omega^2_2\big]\ .
\eqn\gfivea
$$
This metric  as $s\rightarrow +\infty$ (or
$\psi\rightarrow {\pi\over2}$) behaves as
$$
ds^2\sim R^2 dw^2 + {c_1^2 R^2} \big[ (d({x^0\over R c_1})+\cos\theta
d\phi)^2+{1\over
c_1^2}(d\theta^2 +\sin^2\theta d\phi^2)\big ]
\eqn\gfive
$$
where $w=s^{-1}$. Note that, if $c_1=1$, the manifold $\cM$ near
$\psi={\pi\over2}$ becomes ${\cal
R}\times S^3$ and $S^3$ has radius $2 R$ but eventually the metric $g$ will
become singular at
$\cot \psi+1=0$ as we have mentioned above.  The metric $g$ in the
parameterisation \gfivea\
involving the co-ordinates $\{s, x^0, \theta, \psi\}$ is the same as the metric
given by the authors
of ref. [\dv] (up to a gauge choice for the one-form $\omega$, and a
relabelling of some of
parameters of the metric and  some of the co-ordinates of $\cM$) for the
Taub-NUT geometry with
torsion. The same authors had also observed that the space of orbits ${\cal N}$
for this metric is
the three-sphere.  But our conclusion that this metric is singular is opposite
from that of ref.
[\dv].  The torsion is
$$
H=-\lambda c_1 R^3 \sin^2\psi \sin\theta (c_1 \cot\psi+c_0) d\psi d\theta
d\phi\ ,
\eqn\gsix
$$
and it is non-singular. Finally, case (iii) the constant $c_0\not=0$ and the
constant $c_1=0$, in
this case the metric $g$ and the torsion $H$ become those of the $SU(2)\times
U(1)$ WZW model.


\chapter{Concluding Remarks}

We have determined the metric and torsion couplings of (4,0)-supersymmetric
sigma models with a
Noether symmetry generated by a tri-holomorphic Killing vector field and target
space a
four-dimensional manifold.  These metrics are naturally associated to monopoles
on the three-sphere
 which is the space of orbits of the group action generated by the
tri-holomorphic Killing vector
field on the sigma model target manifold. In relation to this result, it is
worth pointing out that
the hyper-K\"ahler Gibbons-Hawking metrics are associated with monopoles on the
flat three-space, and
the scalar flat K\"ahler LeBrun metrics [\leb] are associated with monopoles on
the hyperbolic
three-space.  We have also studied the global properties of a subclass of these
metrics that admit
an isometry group with Lie algebra $u(2)$ acting on the sigma model manifold
with three-dimensional
orbits, for models with non-zero torsion, and we have found that, apart from
the  metric of the WZW
model with target space the group
$SU(2)\times U(1)$, are singular. It also seems likely that all the remaining
metrics, for models
with non-zero torsion, are singular as well.

The class of the above metrics that are singular cannot be thought as
gravitational instantons
associated with some string inspired extension of the Einstein gravity.
However they may serve as
couplings of supersymmetric sigma models way from the singularities.  Such
sigma models will be
ultra-violet finite by the arguments of refs. [\hpa, \hpb].  It may also be
that the (4,0)
supersymmetry of the above sigma models can be extended to a (4,4) one by an
appropriate addition of
a fermionic sector. This suggestion is supported by the fact that the
(4,0)-supersymmetric WZW model
with target space the group $SU(2)\times U(1)$ admits such an extension.

The T-duality transformation can be easily applied to the couplings $g$ and $b$
of the
(4,0)-supersymmetric sigma model with target space a four-dimensional manifold
with respect to
the tri-holomorphic Killing vector field $X$ to give the couplings $g'$, $b'$
and dilaton of the dual
model [\br] (see also [\rg]).  In fact the couplings $g'$ and $b'$ can be
simplified in this case
because,  as we have shown in section 3, $X\cdot H=0$. In addition, the T-dual
theories of the above
(4,0)-supersymmetric sigma models are expected to be ultra-violet finite since
the latter are
ultra-violet finite.  Finally, the application of the new metrics that we have
derived from
consideration of (4,0)-supersymmetric models to string theory and conformal
field theory needs
further study. Such an investigation will involve the construction of the
associated superconformal
field theory.

\vskip 0.5cm

\noindent{\bf Acknowledgments:}  I would like to thank Gary Gibbons and  Paul
Townsend for many
useful discussions on the global properties of four-metrics with singularities,
Chris Hull for his
comments on T-duality and Emery Sokatchev for correspondence. I am supported by
a University Research
Fellowship from the Royal Society.

\refout

\bye